\pdfoutput=1

\documentclass[11pt]{article}

\usepackage{ACL2023}

\usepackage{times}
\usepackage{latexsym}
\usepackage[T1]{fontenc}

\usepackage[utf8]{inputenc}

\usepackage{microtype}

\usepackage{inconsolata}

\usepackage{hyperref}
\usepackage{adjustbox}
        
\usepackage{amsmath,amssymb,amsfonts}
\usepackage[linesnumbered,ruled,vlined]{algorithm2e}
\usepackage{setspace}
\usepackage{algorithmic}
\usepackage{float}
\usepackage{graphicx}
\usepackage{textcomp}
\usepackage{xcolor}
\usepackage{multirow}
\usepackage{mdframed}
\usepackage{balance}
\usepackage{booktabs}
\usepackage{tcolorbox}
\usepackage{makecell}
\usepackage{threeparttable}
\usepackage{colortbl}
\usepackage{subfigure}
\usepackage{hhline}
\usepackage{pifont}
\usepackage{listings}
\usepackage{enumitem}

\newcommand{\parabf}[1]{\noindent\textbf{#1}}

\definecolor{ggray}{HTML}{eff0f0}
\definecolor{gggray}{HTML}{E8E8E8}
\definecolor{ggggray}{HTML}{BEBEBE}
\definecolor{myblue}{RGB}{255,255,255}
\definecolor{myyellow}{HTML}{FFF2CC}

\newcommand{\ie}{{i.e.,}\xspace}
\newcommand{\eg}{{e.g.,}\xspace}

\newcommand{\NULL}{\texttt{NULL}\xspace}

\newcommand{\app}{\textsc{LLM4PFA}}
\newcommand{\appva}{\textsc{NoContext}}
\newcommand{\appvb}{\textsc{BatchSol}}
\newcommand{\ourbench}{\textsc{SAFP-Bench-C}}

\newcommand{\ruleconstraint}{feasible path conditional expression\xspace}
\newcommand{\llmconstraint}{symbolic range\xspace}
\newcommand{\constraint}{feasible path constraints\xspace}
\newcommand{\sourcesym}{ S_{f}\xspace}
\newcommand{\sinksym}{ T_{f} \xspace}
\newcommand{\cdpsym}{N_{cs}(\sourcesym, \sinksym)\xspace}
\newcommand{\cdpcsym}{ FPE(\sourcesym, \sinksym)\xspace}
\newcommand{\cdpoutsinksym}{N_{\text{e}}(\sinksym)\xspace}
\newcommand{\cdpjumpsym}{ N_{\text{jump}} (\sourcesym, \sinksym)\xspace}

\newcommand{\ddpcsym}{ \Sigma_{v}\xspace}
\newcommand{\rv}{ \Sigma_{v}( P)\xspace}
\newcommand{\rf}{ \Sigma_{f}(P)\xspace}

\usepackage[utf8]{inputenc}
\usepackage[english]{babel}

\newcounter{finding}

\newcommand{\distance}{5pt}
\title{Minimizing False Positives in Static Bug Detection via LLM-Enhanced \\Path Feasibility Analysis}

\author{
    Xueying Du\textsuperscript{1} \quad Kai Yu\textsuperscript{1} \quad Chong Wang\textsuperscript{2} \quad Yi Zou\textsuperscript{1} \quad Wentai Deng\textsuperscript{1} \\
    \textbf{Zuoyu Ou\textsuperscript{1} \quad Xin Peng\textsuperscript{1} \quad Lingming Zhang\textsuperscript{3} \quad Yiling Lou\textsuperscript{1}} \\
    \textsuperscript{1}Fudan University, China \\
    \textsuperscript{2} Nanyang Technological University, Singapore \\
    \textsuperscript{3}University of Illinois Urbana-Champaign, USA}

\begin{document}
\maketitle

\begin{abstract}
Static bug analyzers play a crucial role in ensuring software quality. However, existing analyzers for bug detection in large codebases often suffer from high false positive rates. This is primarily due to the limited capabilities of analyzers in path feasibility validation with multiple conditional branches and complex data dependencies. While current LLM-based approaches attempt to address this issue, their effectiveness remains limited due to insufficient constraint cascade analysis and scalability challenges in large projects. 
To address this challenge, we propose an iterative path feasibility analysis framework \app{}. By leveraging LLM agent based targeted constraint reasoning, and key context-aware analysis driven by agent planning, \app{} effectively enhances complex inter-procedural path feasibility analysis for minimizing false positives in static bug detection. 
Evaluation results show that \app{} precisely filters out 72\% to 96\% false positives reported during static bug detection, significantly outperforming all the baselines by 41.1\% - 105.7\% improvements; meanwhile \app{} only misses 3 real bugs of 45 true positives.
\end{abstract}

\section{Introduction}

Static bug analyzers play a crucial role in ensuring software security, reliability, and maintainability. They help identify potential bugs early in the development process by scanning the codebase without extensive program execution~\cite{cai2023place,cai2022peahen,cai2021canary,shi2020conquering,shi2018pinpoint,sui2012static,vassallo2020developers,yan2018spatio}. Popular analyzers like GitHub CodeQL~\cite{codeql}, Facebook Infer~\cite{infer}, and CppCheck~\cite{cppcheck} are widely used for industrial bug detection. Their analysis of many bug types (\eg{} null pointer dereference and use-after-free) typically follows a standardized approach: identifying sources (\ie{} potential origins of issues) and sinks (\ie{} locations where issues manifest) based on predefined rules, then checking whether a feasible execution path exists from sources to sinks. If such a path is found, the tools flag a potential bug.

However, in complex software systems, sources and sinks often reside in different functions. In cases where function call chains between a source and a sink are lengthy and involve multiple conditional branches and complex data dependencies, traditional static analyzers struggle to comprehensively validate path feasibility between sources and sinks, resulting in high false positive rates. For instance, our manual annotation of sampled bug reports reveals that popular tools produce an extremely high rate (e.g., over 90\%) of false positives for use-after-free bugs across large-scale software such as Linux Kernel. This significantly burdens developers due to the large efforts in manually reviewing the false alarms reported by these tools. 

\textbf{Existing Techniques.} As Large Language Models (LLMs) are increasingly applied to code-related tasks~\cite{du2024evaluating,liu2024large,nam2024using, du2024vulragenhancingllmbasedvulnerability}, researchers have leveraged their advanced program understanding capabilities to reduce false positives generated by static analyzers. However, existing LLM-based approaches have limited effectiveness in handling path feasibility analysis within large, dependency-rich codebases. Specifically, LLM4SA~\cite{llm4sa} feeds entire source-to-sink call chains along with bug reports into LLMs at once to identify false positives, without conducting fine-grained constraint analysis and relying entirely on the LLM's ability to assess complex path feasibility. LLMDFA~\cite{llmdfa} performs data flow analysis for each function along all potential execution paths between sources and sinks, which results in significant time and computational cost when applied to large-scale projects with complex call dependencies.
While symbolic execution can perform path feasibility analysis, it suffers from path explosion when traversing all possible paths, particularly when dealing with multiple conditional branches, loops, nested function invocations, or compound data types~\cite{10.1145/3182657}. Therefore, it cannot scale to large and complex software systems in practice. 

\textbf{Our approach. }
To address these limitations, we propose \app{}, an LLM-powered path feasibility analysis framework designed to minimize false positives in static bug detection for complex software projects. 
Specifically, Given a function call trace and a target variable with its source and sink, \app{} iteratively performs two-phase analysis for each function: (1) \textit{Feasibility Constraints Extraction} first identifies critical conditional expressions impacting sink reachability; then employs LLM agents to conduct symbolic range reasoning for variables and function calls within each expression. (2) \textit{Constraints Solving} employs constraint solver to iteratively solve the SMT query scripts generated by LLM. 
Overall, \app{} offers the following advantages compared to traditional symbolic execution and existing LLM-based methods: 

\begin{itemize}[itemsep=2pt,topsep=0pt,parsep=0pt]
\item \textit{Precise complex constraint cascade analysis.} Breaking down inter-procedural feasibility analysis into localized symbolic range reasoning, \app{} amplifies LLMs' program comprehension strength on focused code segments~\cite{li2023loogle}, thereby significantly enhancing reasoning effectiveness.

\item \textit{Efficient complex context analysis.} Integrating prioritized \constraint{} extraction method with agent-guided planning, \app{} effectively mitigates path explosion in  sequentially traversing deeply-nested contextual functions.
\end{itemize}

\textbf{Evaluation.}
We manually construct a new dataset, SAFP-BENCH-C, for a more realistic evaluation, which takes approximately 100 person-hours.  It contains 364 warnings generated by three advanced static analyzers, covering three important bug types (i.e., null pointer dereference, buffer overflow, and use after free) in large-scale software projects, including the Linux Kernel (with over 18 million lines of code), OpenSSL (with over 300K lines of code), and Libav (with over 600K lines of code). Our evaluation shows that \app{} precisely filters out 72\% to 96\%) false positives reported during static bug detection, significantly outperforming all the baselines by 41.1\% - 105.7\% improvements; meanwhile \app{} only misses 3 real bugs of 45 true positives. 
Furthermore, \app{} shows consistent effectiveness across four state-of-the-art LLMs, indicating the generalizability of our approach. Ablation studies further reveal that \app{} outperforms two ablation variants, underscoring the key contribution of LLM agent based self-planning context analysis and iterative constraint solving proposed in our framework.

This paper makes the following contributions:

\begin{itemize}[itemsep=2pt,topsep=0pt,parsep=0pt]

\item \textbf{A novel technique} that effectively enhances  complex inter-procedural feasibility analysis to minimize false positives in bug detection.

\item \textbf{A new dataset} designed for a more realistic evaluation of false positive reduction, with approximately 100 person-hours dedicated to manual construction.

\item \textbf{An extensive evaluation} demonstrating the effectiveness of \app{} in filtering out false positives and identifying real bugs. 
\end{itemize}

\begin{figure*}[htb] 
\centering        
\subfigure[Complex Constraint Cascades] 
{    
\begin{minipage}[t]{0.33\linewidth}
\centering 
\includegraphics[width=\linewidth]{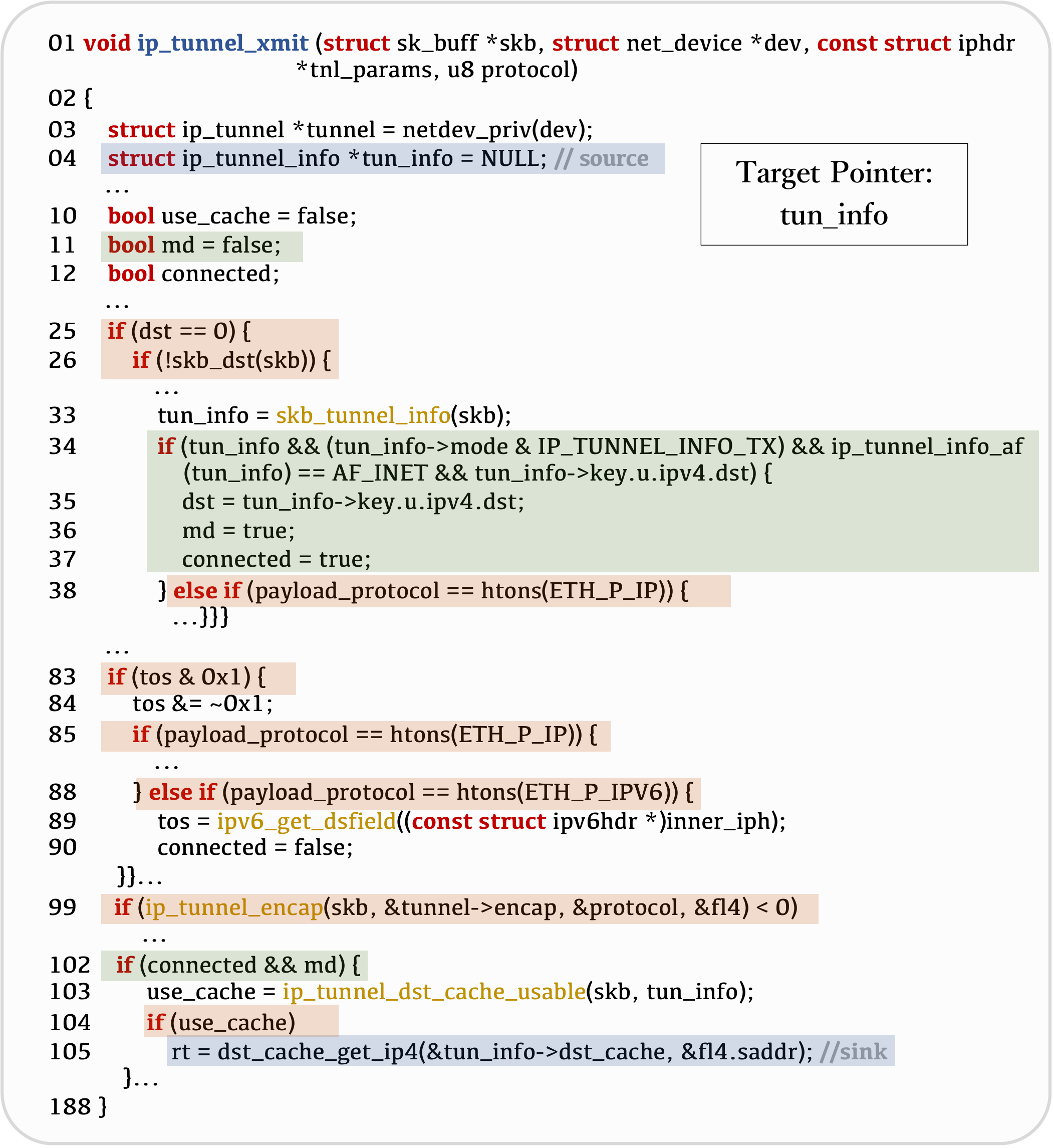}    \label{fig:example2}
\end{minipage}
}    
\subfigure[Complex Constraint-Related Contextual Analysis] 
{  
\begin{minipage}[t]{0.64\linewidth}
\centering   
\includegraphics[width=\linewidth]{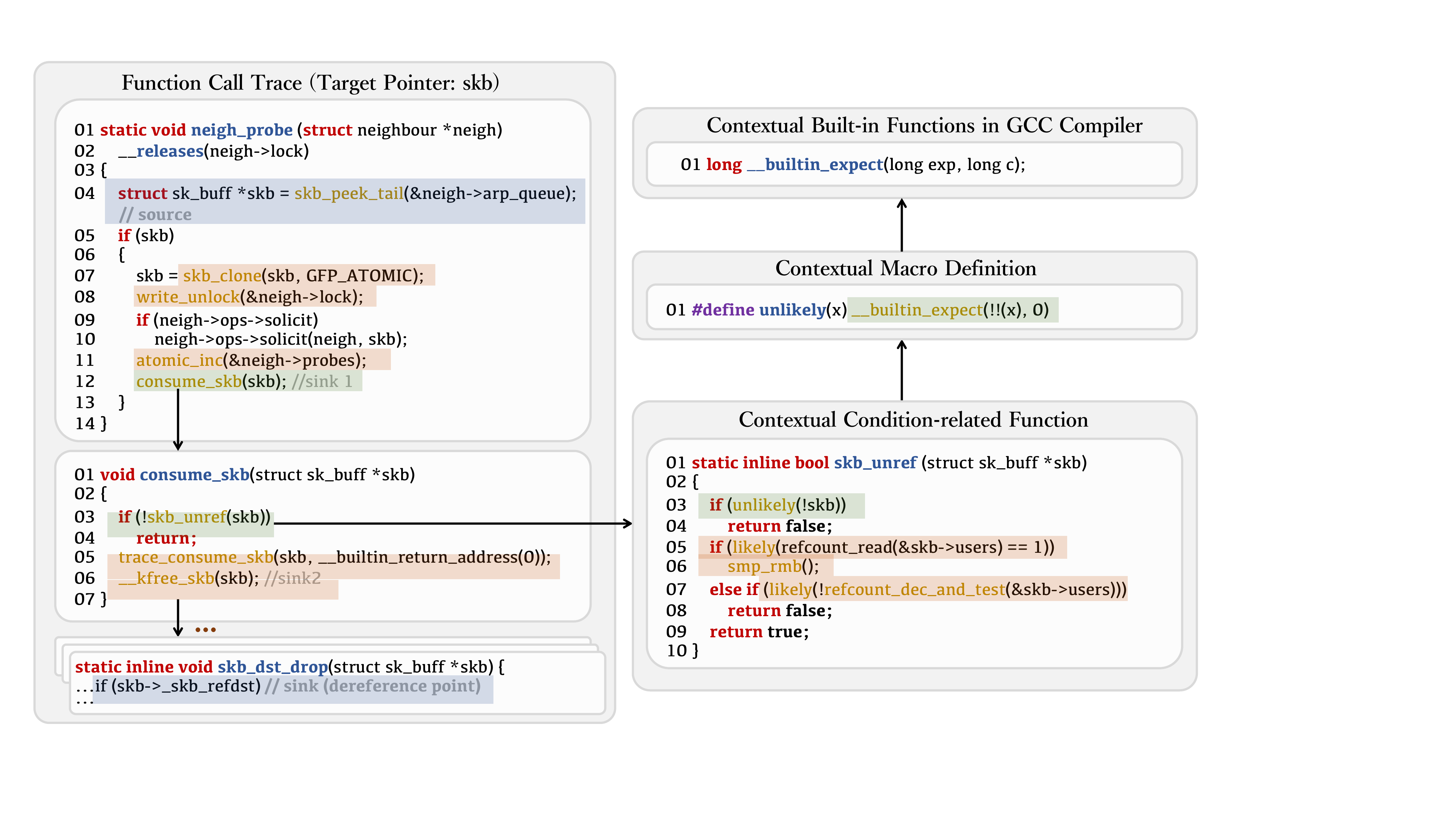}  
\label{fig:example1}
\end{minipage}   
}   
\caption{Motivating Examples}
\label{fig:exmaples}        
\end{figure*}

\section{Preliminaries \& Motivation}
This section introduces key preliminaries of static bug detection and presents motivating examples to illustrate the false positive issues related to path feasibility validation.

\subsection{Static Bug Detection}
Many bugs, such as Null Pointer Dereference (NPD) and Use After Free (UAF),  follow a common pattern where a \textit{target variable} feasibly flows from a \textit{source} (where a potential issue originates) to a \textit{sink} (where the issue manifests). 

\textbf{Source and Sink Identification.}
The specific definitions of the target variable, source, and sink vary by bug type. For instance, in NPD (Null Pointer Dereference) bugs, the target variable is a pointer (\eg \textit{tun\_info} in Figure~\ref{fig:example2}), the source is the code expression that defines or assigns it as \NULL (\eg Line 4 in Figure~\ref{fig:example2}), and the sink is the code expression that dereferences the pointer (e.g., Line 105 in Figure~\ref{fig:example2}). 
For a specific bug type $bt$, given a code project $\mathcal{P}$ consisting of multiple code files and lines, the identification of the sources and sinks associated with the target variables can typically be formalized through the procedures:
$$ \textsc{FindSource}^{bt}: \Sigma_\mathcal{P} \to \Sigma_{var} \times \Sigma_{expr} $$
$$ \textsc{FindSink}^{bt}: \Sigma_\mathcal{P} \to \Sigma_{var} \times \Sigma_{expr}, $$
where $\Sigma_{(*)}$ represents the domain of $(*)$, \ie $\Sigma_{var}$ and $\Sigma_{expr}$ denote the complete value domains of variables and code expressions, respectively. Each $source$ or $sink$ is represented as a pair $(var, expr)$, consisting of a target variable $var$ and its corresponding code expression $expr$.

\textbf{Data Flow Analysis.}
For each source-sink pair $(source, sink)$ sharing the same target variable $var$, data flow analysis aims to determine whether $var$ can flow from $source$ to $sink$. If true, a corresponding bug is reported. For instance, in Figure~\ref{fig:example2}, data flow analysis for the pointer \textit{tun\_info} is performed between Line 4 (source) and Line 105 (sink). If a feasible execution path is found, an NPD bug is reported at the sink. This analysis can be formalized as the procedure:   
$$ \textsc{TraceFlow}: \Sigma_{source} \times \Sigma_{sink} \to \{\textit{TRUE}, \textit{FALSE}\}, $$  
which returns \textit{TRUE} if any feasible execution path exists between $source$ and $sink$, and \textit{FALSE} otherwise.

\textbf{Path Feasibility Validation.}
Performing precise data flow analysis is challenging due to the complexity of validating path feasibility. Sources and sinks often reside in different functions, necessitating inter-procedural analysis to trace all potential execution paths. When the function call chain between a source and a sink is lengthy and includes multiple conditional branches, nested function invocations, or compound data types, the complexity of path feasibility validation
increases exponentially (We will illustrate the challenges with examples in Section~\ref{sec:example}).  
However, traditional static analysis tools like CodeQL and Infer often incorporate over-approximation during contextual analysis, so as to reduce the complexity of path feasibility validation. As a result, there are high false positive rates in their reported bug alarms in practice.

\subsection{Motivating Examples}
\label{sec:example}
We present two examples to further illustrate the challenges of validating path feasibility, which lead to false positives in the NPD bug reported by CodeQL on the Linux kernel. 

\textbf{Challenge-1: Complex Constraint Cascades in Path Tracing.} 
Figure~\ref{fig:example2} presents a false positive reported by CodeQL. In this example,  the green-highlighted sections involve value assignments and constraint validation for the two \textit{bool} variables \textit{connected} and \textit{md}. According to the constraint in Line 34, we can deduce that only when the target pointer \textit{tun\_info} is \textbf{not} \NULL, both \textit{connected} and \textit{md} are \textit{TRUE} (Lines 36 and 37). Based on the constraint in Line 102, only if both \textit{connected} and \textit{md} are \textit{TRUE}, can the dereference operation (the sink) at Line 105 potentially be executed. According to both constraints, it is clear that no feasible path exists to reach the sink, and thus no NPD occurs. Beyond these green-highlighted sections, the example also involves numerous preceding and succeeding conditional constraints (highlighted in orange). Only by comprehensively cascading these constraints and analyzing the potential execution paths,  can accurate path feasibility validation be achieved and false positives be eliminated. Such a high complexity is prevalent in real-world software systems, which often involve intricate function call chains and deeply nested conditional structures, making path feasibility analysis even  challenging.

\textbf{Challenge-2: Contextual Analysis.} 
In addition to the source-to-sink function call trace, precise path feasibility analysis requires reasoning numerous variable values and function call return values, further complicating the analysis process. Figure~\ref{fig:example1} illustrates a false positive resulting from complex contextual analysis. In Line 3 of the second function in the trace, \textit{consume\_skb} correctly performs a null check on the target pointer \textit{skb} by invoking \textit{skb\_unref}, which returns \textit{false} when \textit{skb} is null. This causes the program to \textit{return} early, preventing the null pointer from propagating to the sink. Notably, Figure~\ref{fig:example1} only reveals the minimal contextual information (highlighted in green) necessary for the manual inspection to identify the false positive. Existing LLM-based approaches would consider a broader set of contextual function calls (highlighted in orange), significantly increasing the complexity of context-sensitive analysis. Furthermore, due to the intricacies of the Linux kernel, this analysis incorporates not only function calls but also macro definitions and built-in GCC compiler functions, presenting additional challenges for traditional symbolic execution techniques.

\begin{figure*}[htb]
	\centering
	\includegraphics[width=0.9\linewidth]{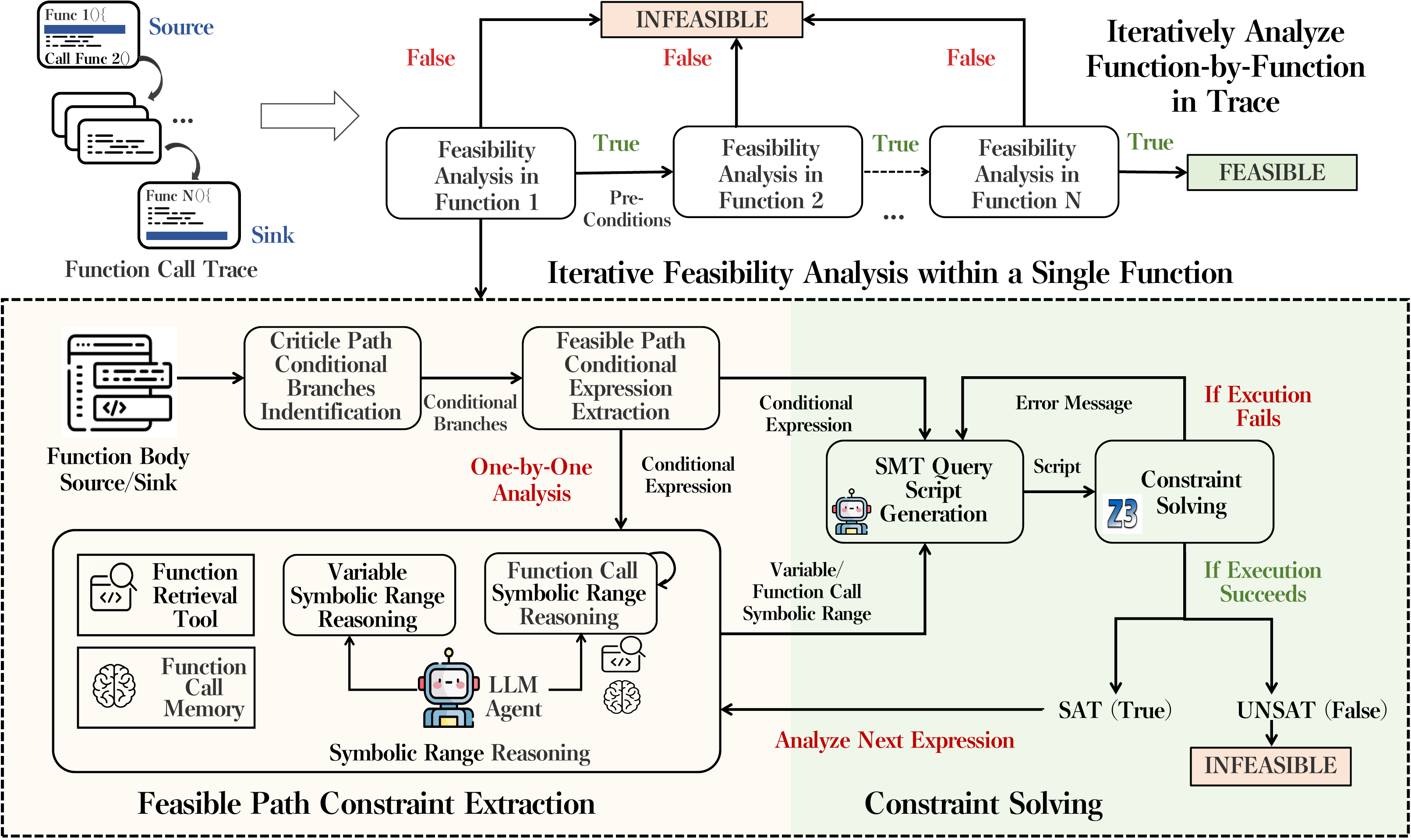}
	\caption{Overview of \app{}}
	\label{fig:overview}
 
\end{figure*}

\section{Approach}
\subsection{Overview}

In this work, we present an iterative path feasibility validation framework \app{} to enhance complex inter-procedural feasibility analysis for minimizing false positives in static bug detection.

Figure~\ref{fig:overview} shows the overview of our approach. Given a target variable, its source and sink, and the function call trace from source to sink, \app{} performs  iterative feasibility analysis function-by-function. 
Each function is analyzed iteratively by collecting \constraint{} and solving them to determine whether a feasible source-to-sink path exists. The process returns \textit{True} if any feasible path exists, and \textit{False} otherwise. This iterative process continues until either the analysis of all functions in the call trace is complete, or a  \textit{False} result is encountered.

The detailed iterative analysis within each function proceeds as follows:

\begin{itemize}[itemsep=2pt,topsep=0pt,parsep=0pt]

\item \textbf{Feasibility Constraints Extraction (Section~\ref{sec:path_constraints_extraction})}: Based on the Control-Flow Graph (CFG), \app{} first identifies the critical path conditional branches between source and sink, and summarizes \ruleconstraint{s}. Subsequently, \app{} incorporates LLM agents to conduct symbolic range reasoning for variables and function calls within each expression individually. 

\item \textbf{Constraints Solving (Section~\ref{sec:constraints_solving})}: \app{} leverages LLM to convert all extracted \constraint{} into SMT query scripts, and invokes the constraint solver (i.e., Z3) for constraint  solving.

\end{itemize}

\subsection{Feasibility Constraints Extraction}\label{sec:path_constraints_extraction}

Given a function 
$f$ with the CFG entry point
$\sourcesym{}$ and the exit point
$\sinksym{}$, \app{} first extracts the \constraint{} from $\sourcesym{}$ to $\sinksym{}$. 

To improve precision and avoid the inefficiency of exhaustively traversing all possible paths, we begin by extracting a set of critical path conditional expressions that directly impact the reachability of $\sinksym $ (Section~\ref{sec:constraints_extraction}). Subsequently, LLM agent reasoning is employed to determine the symbolic ranges of variables and function call return values within these expressions (Section~\ref{sec:constraints_complementation}). These conditional expressions and symbolic ranges collectively form the \constraint{}.

\subsubsection{Critical Path Conditional Expression Extraction}\label{sec:constraints_extraction}

Based on the extracted CFG of $f$, \app{} first identifies the critical path conditional
branches, which are a set of conditional statement nodes (i.e., branching and loop constructs) that directly impact the reachability of $\sinksym $ at the control flow level. These branches are represented as:
$$\cdpsym{} = \cdpoutsinksym{} \cup \cdpjumpsym{}$$
where $\cdpoutsinksym{}$ represents all conditional statement nodes nested outside the basic block containing $\sinksym{}$; and
$\cdpjumpsym{}$ refers to conditional statement nodes nested outside any basic block containing \textit{Jump/Return} statements encountered between $\sourcesym{}$ and $\sinksym{}$.
For branching statements, we focus on \textit{If} statements (\ie{} \textit{if}, \textit{else}, \textit{else if} and \textit{switch}); for \textit{Loop} statements, we focus on \textit{while} and \textit{for} statements. Regarding \textit{Jump/Return} statements, we consider all \textit{goto} and \textit{return} statements, as well as \textit{break} and \textit{continue} statements within loop constructs that are nested outside $\sinksym{}$.

Based on these critical path conditional branches, \app{} further extracts  the \ruleconstraint{s}, which are  represented as:

$$\cdpcsym{} = \left\{ expr_{f} \mid expr_{f} \in \cdpoutsinksym{} \right\} \nonumber $$ 
$$\quad \cup \left\{ \neg expr_{if} \mid expr_{if} \in \cdpjumpsym{} \right\}$$
Where a feasible path exists only when all $expr_{f}$ are satisfied (i.e., the conditions for reaching the exit point branch are fulfilled), and none of $expr_{if}$ is  fully satisfied (i.e., there is no early jump to other branches).

\subsubsection{Context-Aware Symbolic Range Reasoning}
\label{sec:constraints_complementation}

In this step, \app{} utilizes LLM agents to reason all \llmconstraint{s} $\ddpcsym$ of each \ruleconstraint{} one by one. 
Notably, $\ddpcsym{}$ encompasses the \llmconstraint{s} from both variables and the function call return values in conditional expressions. These constraint-related function calls are not part of the input function call trace, and thus require additional contextual analysis.

\parabf{Variable Symbolic Range Reasoning.} For each variable $v$ in a conditional expression, \app{} reasons the possible \llmconstraint{s} $\rv{}$ of $v$ when the conditional statement is executed. In this context, $P$ denotes the required initial states of the program.

The initial states $P$ are the post-conditions following the execution of $\sourcesym{}$. These conditions encompass both the pre-conditions required by $\sourcesym{}$ and the post-conditions of the $\sourcesym{}$ statement. For the initial function in the input trace, $P$ serves directly as the post-condition of $\sourcesym{}$. For subsequent functions, $P$ further incorporates all \constraint{} associated with parameters passed to the current function, as extracted from the preceding function. This ensures the propagation of constraints through function call trace, enhancing the rigor of inter-procedural analysis.
The definition of $P$ is adapted based on the specific bug type.  In the case of NPD detection, the \textit{NULL} state of the target pointer is propagated as $P$ throughout the entire function call trace analysis.

Inspired by chain-of-thought (CoT), \app{} directs the LLM agent to perform a step-by-step simulated execution of function code, enabling more precise reasoning by LLM. Furthermore, example analyses are included within the prompt to facilitate few-shot learning for LLM agents. The detailed prompt design is in Appendix \ref{ap:prompt}.

\parabf{Iterative constraint-related function call Symbolic range reasoning.} 
For constraint-related function calls $fc$ within a conditional expression, 
\app{} aims to determine the \llmconstraint{} $\rf$ of the function return value under the required initial states $P$. The definitions of $P$ align with $\rv{}$.

\begin{figure}[htb]
	\centering
	\includegraphics[width=1\linewidth]{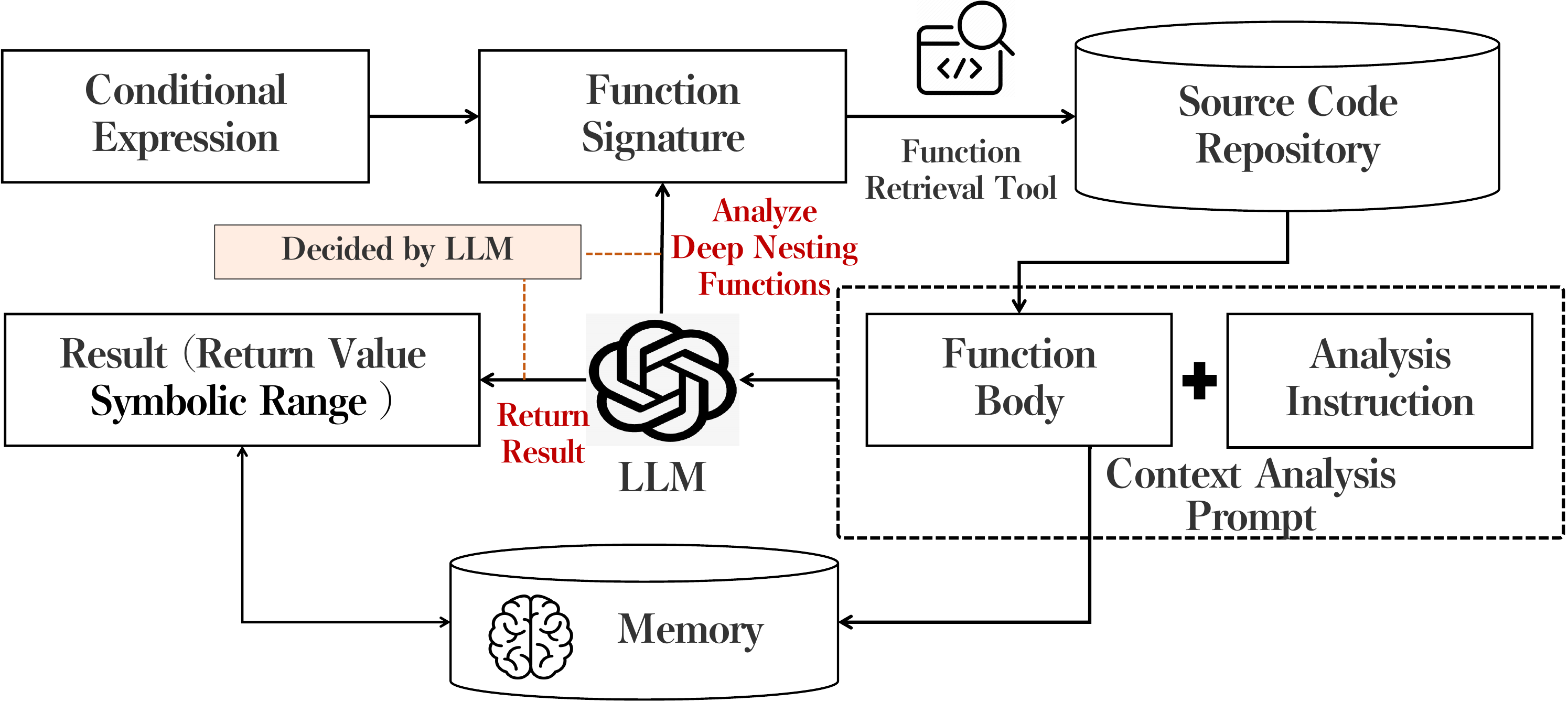}
	\caption{LLM Agent-based Context-aware Return Value Range Analysis of Function Calls}
	\label{fig:context_analysis}
 
\end{figure}

Figure \ref{fig:context_analysis} illustrates the framework for reasoning \llmconstraint{s} of constraint-related function call return values. This reasoning process may require deep multi-layered contextual function analysis. 
To address the challenges posed by the complexity of context analysis (Challenge-2), we propose an LLM agent-based framework that leverages the tool-usage and self-planning capabilities of LLM agents for precise context analysis iteratively.  

Given that \ruleconstraint{s} may involve multiple function calls, only those with parameters relevant to $P$ are selected for \llmconstraint{} reasoning.  
For each contextual function under analysis, \app{} retrieves its function body from the source code repository using a function retrieval tool. This function body, along with the initial states $P$, is then provided to LLMs  for reasoning. 
During the reasoning process, the LLM agent dynamically determines whether additional deeper contextual functions are needed to refine the analysis, leveraging its self-planning capabilities
. Contextual functions that are identified as affecting the return values are further analyzed with  the same process. This iteration continues until the LLM agent finalizes the result or reaches the predefined maximum iteration limit (5 in this work).

To minimize resource consumption due to redundant analyses of identical function calls, we introduce a memory module. This module stores the return value \llmconstraint{} of previously analyzed function calls under distinct $P$. In subsequent analyses, results for identical scenarios can be retrieved directly from the memory module.

\subsection{Constraints Solving}\label{sec:constraints_solving}

Given the \constraint{} for $\sinksym{}$, \app{} iteratively employs LLM to convert them into SMT query scripts, and invokes Z3 solver for constraint solving.

\begin{figure}[htb]
	\centering
	\includegraphics[width=0.9\linewidth]{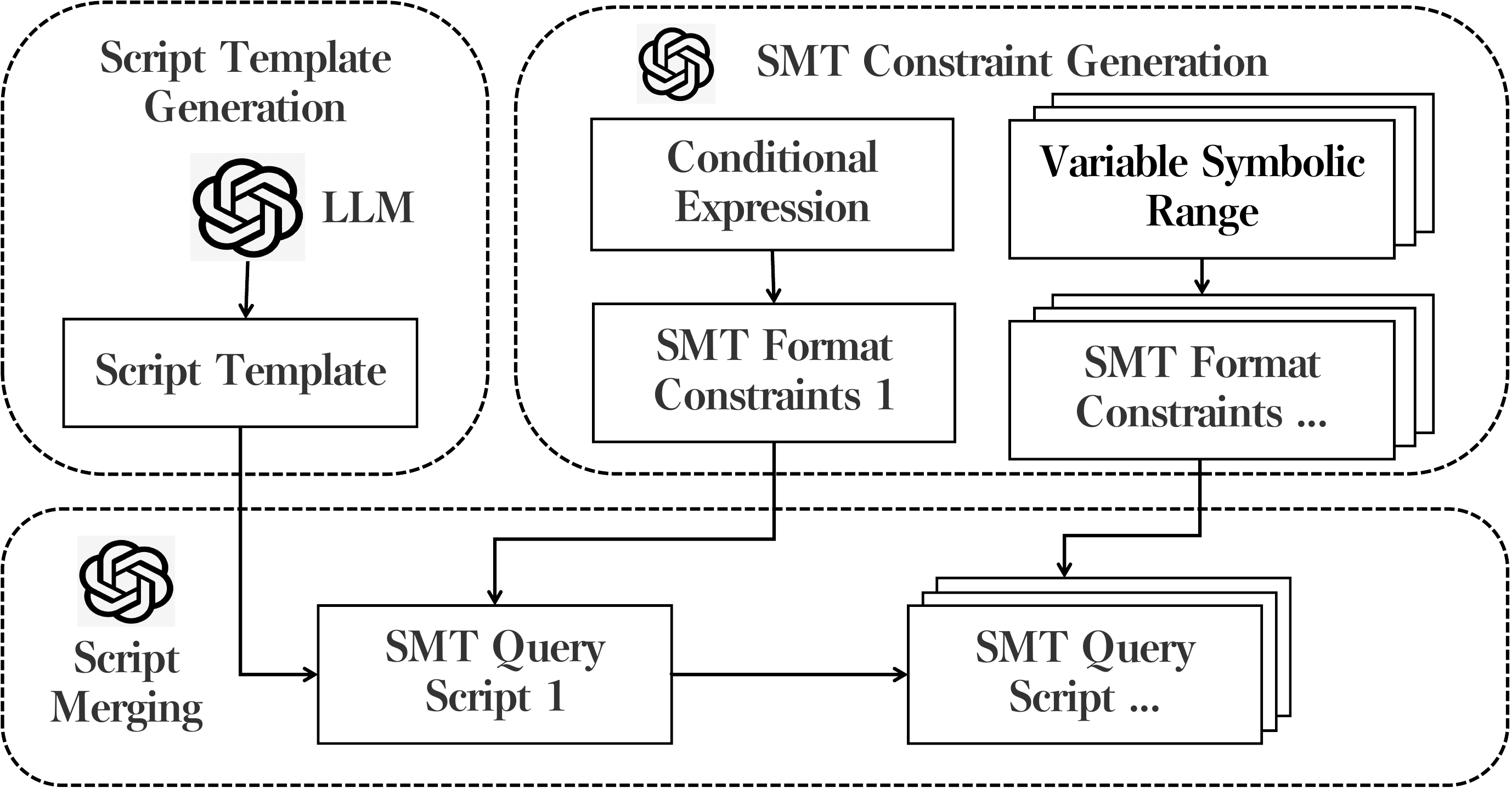}
	\caption{SMT Query Script Generation}
	\label{fig:solving}
 
\end{figure}

When analyzing complex functions, \app{} may solve numerous \constraint{}. Directly converting all constraints into SMT query scripts at once can overwhelm the LLM. To address this, we propose an iterative generation strategy.
Figure \ref{fig:solving} illustrates this generation process, which consists of three key steps:

\begin{itemize}[itemsep=2pt,topsep=0pt,parsep=0pt]

\item \textbf{Script template generation.} We utilize LLM to generate a Python template of the SMT query script, including imported packages and constraint solver definitions.

\item \textbf{SMT constraints generation.} With few-shot examples, LLMs then convert  path constraints into SMT constraints (i.e., \textit{z3.add()} format).

\item \textbf{Script merging.} Lastly, LLMs merge SMT constraints with the template to create the final script, ensuring consistency between SMT variable definitions and SMT constraints.

\end{itemize}

After generating the initial script (Script 1 in Fig~\ref{fig:solving}), the first round of constraint solving is performed. If the solver returns an \textit{UNSAT} (\ie infeasible) result, further \llmconstraint{} reasoning is terminated. During this process, each set of \llmconstraint{s} in conditional expressions inferred by the LLM will be converted into SMT constraints and immediately merged into the script for constraint solving. Once the solver returns \textit{UNSAT}, further reasoning is halted to avoid inefficient analysis of additional constraints.

To address script generation errors, \app{} incorporates error messages from the Z3 solver into the prompt, enabling targeted fixes by the LLM. This process repeats until the Z3 solver successfully resolves the constraints or the maximum number of fix attempts (3  rounds in our setting) is reached. The detailed prompt design of SMT query script generation is in Appendix~\ref{ap:prompt}.

\section{Evaluation}

\subsection{Dataset}

We first manually label and construct a new benchmark \ourbench{}, which includes false positive reported by state-of-the-art static analyzers on real-world complex software systems.

\textbf{Bug Types.} We target three representative categories of critical bugs: null pointer dereference (NPD), use-after-free (UAF), and buffer overflow (BOF). 
\textbf{Target Projects.} We select the latest versions of three well-maintained and large-scale C/C++ open-source projects for scanning, including the Linux kernel~\cite{kernel}, OpenSSL~\cite{openssl}, and Libav~\cite{libav}.
\textbf{Static Analyzers.} We use three state-of-the-art static analyzers: CodeQL~\cite{codeql}, Infer~\cite{infer}, and CppCheck~\cite{cppcheck}.

We adopt the following sampling strategy for manual inspection: if a static analyzer detects more than 100 warnings in a project, we randomly sample 100 instances; otherwise, all instances are included. For each warning, two participants will individually label it as True or False. In total, the construction process takes around 100 person-hours. The final benchmark \ourbench{} contains 364 warnings, of which 45 are real bugs. Detailed benchmark statistics are presented in Table \ref{table:benchmark} in Appendix~\ref{ap:dataset} and the detailed labeling process is in Appendix~\ref{ap:dataset}. 
\subsection{Method Comparison}

\parabf{Baselines.} We compare \app{} against two state-of-the-art LLM-based false positive detection methods: LLM4SA~\cite{llm4sa} and LLMDFA~\cite{llmdfa}. 
The detailed baseline settings are in Appendix \ref{ap:baseline}.

\parabf{Setup and Metrics. }We use three commonly used metrics in vulnerability detection tasks: accuracy, precision, and recall. Additionally, we introduce two metrics to evaluate the effectiveness in identifying false positives generated by static analyzers: False Positive Reduction Precision (FPR\_P) and False Positive Reduction Recall (FPR\_R), which specifically measure the precision and recall for negative samples (i.e., false positives) in the dataset.
In this section, for both \app{} and the baselines, we use GPT-4o-mini~\cite{gpt4o} as the backbone LLM. To minimize randomness, we set the temperature to 0, ensuring the LLM performs greedy decoding without sampling.

\parabf{Results.}
Table~\ref{table:rq1} compares \app{} with two baselines on \ourbench{} across three bug categories. 
Overall, \app{} significantly outperforms all baselines across all metrics. Furthermore, \app{} eliminates more false positives than the baselines (with a FPR\_R improvement of 41.1\% - 105.7\%). Particularly, \app{} precisely filters out 72\% to 96\% false alarms on different bugs. Meanwhile, \app{} still detected 42 real bugs (true positives) out of 45, achieving a high recall rate of 0.93 (with a 3.3\% - 38.8\% increase). 
Among the three bug types, the improvement for UFA detection is limited due to the relatively simple nature of the cases, where false positives arise from significant validity checks before use or reallocation after free, which do not involve complex context or path constraints. 

Table~\ref{table:rq1_2} presents the effectiveness of \app{} with three static analyzers in NPD detection. 
Among three analyzers, \app{} demonstrates the greatest advantage in eliminating false positives from CodeQL (with a FPR\_R improvement of 97.2\% - 129.0\%). This improvement is attributed to CodeQL's relatively comprehensive null pointer check rules, which lead to more complex false positives, such as the complex constraint cascade (\eg{} Example-1 in Section~\ref{sec:example}), and requires more extensive context analysis. Specifically, 44\% of CodeQL cases involve context analysis, whereas only 0.29\% and 0.36\% of cases from cppcheck and infer involve context analysis, respectively.

\begin{table}[htb]
    \centering
    \caption{Comparison on Different Bugs}\label{table:rq1}
    \footnotesize
    \begin{adjustbox}{width=1\linewidth}
        \begin{tabular}{c|c|c|c|c|c|c}
        \hline
        \textbf{Type} & \textbf{Tech.} & \textbf{Acc.} &
        \textbf{FPR\_P} & \textbf{FPR\_R} &
        \textbf{Pre.} & \textbf{Recall} \\ \hline

        \multirow{4}{*}{NPD}& \app{} & \textbf{0.75}  & \textbf{0.98} & \textbf{0.72} &\textbf{ 0.36} &  \textbf{0.93}\\
        & LLM4SA &  0.57 & 0.97 &  0.51& 0.24 & 0.90\\
        & LLMDFA & 0.39  & 0.88  & 0.35  & 0.13 & 0.67 \\\hline 
        
        \multirow{4}{*}{BOF}& \app{} & \textbf{0.86} & \textbf{1.0} & \textbf{0.86} & / & /  \\
        & LLM4SA & 0.26 & \textbf{1.0} & 0.26 & / & / \\
        & LLMDFA & 0.02  & \textbf{1.0} & 0.02 & / & /\\\hline 

        \multirow{4}{*}{UAF}& \app{} & \textbf{0.96}  & \textbf{1.0} & \textbf{0.96} & / & / \\
        & LLM4SA & 0.92  & \textbf{1.0} & 0.92 & / & / \\
        & LLMDFA & 0.32  & \textbf{1.0} & 0.32 & / & / \\\hline 

        \end{tabular}
    \end{adjustbox}
\end{table}

\begin{table}[htb]
    \centering
    \caption{Comparison on Different Analyzers}\label{table:rq1_2}
    \footnotesize
    \begin{adjustbox}{width=1\linewidth}
        \begin{tabular}{c|c|c|c|c|c|c}
        \hline
        \textbf{Analyzer} & \textbf{Tech.} & \textbf{Acc.} &
        \textbf{FPR\_P} & \textbf{FPR\_R} &
        \textbf{Pre.} & \textbf{Recall} \\ \hline

        \multirow{3}{*}{CodeQL}& \app{} & \textbf{0.75}  & \textbf{0.97} & \textbf{0.71} &\textbf{0.42} &\textbf{0.91}  \\
        & LLM4SA & 0.45  & 0.93 &  0.36& 0.24&0.88\\
        & LLMDFA & 0.38  & 0.85 & 0.31 & 0.17& 0.73\\\hline 
        
        \multirow{3}{*}{Infer}& \app{} & \textbf{0.68 } &\textbf{1.0}  &  \textbf{0.63}& \textbf{0.30}&\textbf{1.0}  \\
        & LLM4SA & 0.59  & \textbf{1.0} & 0.53 & 0.25&\textbf{1.0}\\
        & LLMDFA & 0.27  & 0.71 & 0.26 & 0.07 & 0.33 \\\hline 

        \multirow{3}{*}{Cppcheck}& \app{} & \textbf{0.79}  & \textbf{1.0} & \textbf{0.77} & 0.22 & \textbf{1.0} \\
        & LLM4SA & \textbf{0.79}  & 0.95 &  \textbf{0.77 }& 0.24 & 0.60 \\
        & LLMDFA & 0.43  & 0.93 & 0.43 & \textbf{0.54} & 0.50\\\hline 

        \end{tabular}
    \end{adjustbox}
\end{table}
\subsection{Model Versatility}
To study the generality of \app{} on different backbone LLMs, we build \app{} with four state-of-the-art LLMs that have been widely used in bug detection, including two closed-source models, i.e., GPT-4o-mini~\cite{gpt4o}, Claude Sonnet 3.5~\cite{claude}, and two open-source models, i.e., Qwen2.5-Coder-32B-Instruct~\cite{qwen}, DeepSeek-V2-Instruct~\cite{deepseek}.  
Table~\ref{table:rq2} presents the effectiveness of \app{} across four different LLMs. Overall, \app{} demonstrates significant effectiveness on all models (all outperforming baselines using GPT-4o-mini), showcasing the generalizability of our approach across different LLMs. 

\begin{table}[htb]
    \centering
    \caption{Generality with Different LLMs}\label{table:rq2}
    \footnotesize
    \begin{adjustbox}{width=1\linewidth}
        \begin{tabular}{c|c|c|c|c|c|c}
        \hline
        \textbf{Type} & \textbf{Model} & \textbf{Acc.} &
        \textbf{FPR\_P} & \textbf{FPR\_R} &
        \textbf{Pre.} & \textbf{Recall} \\ \hline

        \multirow{4}{*}{NPD}& GPT-4o-mini &  \textbf{0.75} & \textbf{0.98} & \textbf{0.72 }&\textbf{0.36} & \textbf{0.93} \\
        & Claude & 0.71  & \textbf{0.98} &  0.68 &0.32 &0.90\\
        & Qwen & 0.65  & 0.96 & 0.62 & 0.27& 0.83\\
        & Deepseek & 0.72  & 0.97 & 0.69 & 0.33& 0.88\\ \hline 
        
        \multirow{4}{*}{BOF}& GPT-4o-mini & 0.86 & 1.0 & 0.86 & / & /  \\
        & Claude & 0.86 & 1.0 & 0.86 & / & /  \\
        & Qwen &  0.86 & 1.0 & 0.86 & / & /  \\
        & Deepseek & 0.86 & 1.0 & 0.86 & / & /  \\ \hline 

        \multirow{4}{*}{UAF}& GPT-4o-mini & 0.96  & 1.0 & 0.96 & /& / \\
        & Claude &  0.96  & 1.0 & 0.96 & /& / \\
        & Qwen &  0.96  & 1.0 & 0.96 & /& / \\
        & Deepseek &  0.96  & 1.0 & 0.96 & /& / \\ \hline 

        \end{tabular}
    \end{adjustbox}
\end{table}

\subsection{Ablation Studies}

We evaluate the effectiveness of our iterative refinement-based technical design by comparing \app{} with two ablation variants: 
\appva{} performs \llmconstraint{} reasoning without iterative context analysis driven by the LLM agent;
\appvb{} converting all \constraint{} into a single SMT query script at once for direct constraint solving, without intermediate iterations.

\parabf{Results.}
Table~\ref{table:rq3} presents the comparison results of ablation variants with  gpt-4o-mini. Overall, \app{} demonstrates a significant superiority over \appva{} and \appvb{} in false positive elimination, showing 136.7\% - 173.1\% improvements in FPR\_R. While \appvb{} achieves a recall of 0.95, compared to \app{}'s 0.90 in Linux kernel vulnerability detection, this discrepancy arises from a real bug being misclassified as a false alarm by \app{} due to inaccurate extraction of critical path conditional expressions. This case is still considered a bug alarm in \appvb{} due to a constraint solving failure. A high rate of constraint solving failures in \appvb{} contributes to its low FPR\_R (0.38,  57.8\% less than \app{}). This highlights the key contribution of LLM-based self-planning context analysis and iterative constraint solving in \app{}.

\begin{table}[htb]
    \centering
    \caption{Ablation of \app{}}\label{table:rq3}
    \footnotesize
    \begin{adjustbox}{width=1\linewidth}
        \begin{tabular}{c|c|c|c|c|c|c}
        \hline
        
        \textbf{Project} &\textbf{Tech.} & \textbf{Acc.} &
        \textbf{FPR\_P} & \textbf{FPR\_R} &
        \textbf{Pre.} & \textbf{Recall} \\ \hline
        \multirow{3}{*}{Kernel}
        & \app{} & \textbf{0.9}  & \textbf{0.96} & \textbf{0.90} & \textbf{0.8} & 0.90  \\
        & \appva{} & 0.44  & 0.8  & 0.25 & 0.35 & 0.86  \\
        & \appvb{} & 0.56 & 0.95 & 0.38 & 0.41 & \textbf{0.95} \\\hline 

        \multirow{3}{*}{Openssl}
        & \app{} & \textbf{0.8}  & \textbf{0.97} & \textbf{0.78} & \textbf{0.47} & \textbf{0.89}  \\
        & \appva{} & 0.46  & 0.89  & 0.40 & 0.22 & 0.78  \\
        & \appvb{} & 0.44 & 0.93 & 0.34 & 0.23 & \textbf{0.89} \\\hline 

        \multirow{3}{*}{Libav}
        & \app{} & \textbf{0.51}  & \textbf{1.0} & \textbf{0.49} & \textbf{0.07} & \textbf{1.0 } \\
        & \appva{} & 0.21  & \textbf{1.0 } & 0.18 & 0.04 & \textbf{1.0 } \\
        & \appvb{} & 0.23 & \textbf{1.0} & 0.20 & 0.04 & \textbf{1.0} \\\hline

        \multirow{3}{*}{Total}
        & \app{} & \textbf{0.75} & \textbf{0.97}&\textbf{0.71} &\textbf{0.42}  & 0.91 \\
        & \appva{} & 0.37  & 0.88 & 0.26 & 0.21 &0.85 \\
        & \appvb{} &  0.42& 0.96 & 0.30 & 0.23 &\textbf{0.94} \\\hline

        \end{tabular}
    \end{adjustbox}
\end{table}

\section{Related Work}

\parabf{Integrating LLMs with Static Analysis for Bug Detection.}
Currently, substantial efforts have been made to combine LLMs with static analysis to enhance detection capabilities. Li et al.\cite{li2023hitchhikers} propose a framework that integrates UBITect with LLMs to detect Use-Before-Initialization (UBI) bugs in the Linux kernel; Sun et al.~\cite{sun2023gpt} employ LLMs to assist static analysis in identifying logic vulnerabilities in smart contracts. Li et al.~\cite{li2024llm} propose IRIS, which systematically combines LLMs with the static analysis tool CodeQL to conduct the first whole-repository reasoning approach for comprehensive vulnerability detection. Wen et al.~\cite{llm4sa} present LLM4SA, which utilizes LLMs to process bug warning reports from various static analysis tools, leveraging domain knowledge of LLMs to reduce false positives.
Wang et al.~\cite{wang2023boosting} propose InferROI, which employs LLMs to infer resource-oriented intentions (e.g., acquisition, release, and reachability validation) in code, combined with a two-stage static analysis approach to inspect control-flow paths for potential resource leak detection.

\parabf{LLMs for Program Analysis.}
Shahandashti et al.~\cite{shahandashti2024program} investigate the application of LLMs in both static and dynamic program slicing, utilizing advanced prompting techniques such as few-shot learning and chain-of-thought reasoning.
Wang et al.~\cite{llmdfa} introduce LLMDFA, the first LLM-powered dataflow analysis framework, which decomposes the problem into three components: source/sink extraction, dataflow summarization, and path feasibility validation.
Subsequently, they further propose LLMSAN~\cite{wang2024sanitizing}, which leverages few-shot chain-of-thought prompting to guide LLMs in emitting data-flow paths, with validation performed through program-property decomposition.

To the best of our knowledge, no prior studies have performed the iterative decomposition of complex program analysis tasks or fine-grained symbolic range reasoning. Moreover, our approach utilizes the self-planning capabilities of LLM agents to automate the iterative analysis of deeply nested functions. This effectively mitigates the path explosion problem inherent in complex inter-procedural path feasibility analysis, which existing studies have struggled to address efficiently.

\section{Conclusion}
This paper proposes \app{}, an iterative path feasibility analysis framework powered by LLM agents, which effectively enhances  complex inter-procedural feasibility analysis for minimizing false positives in bug detection. Our evaluation shows that \app{} substantially outperforms all baselines in both bug detection (with 0.93 recall) and false positive elimination. 

\bibliography{ref,custom}
\bibliographystyle{acl_natbib}

\appendix

\section{Dataset }
\label{ap:dataset}


\ourbench{} contains 364 warnings generated by three static analyzers scanning three C/C++ projects, of which 45 are true bugs. Table~\ref{table:benchmark} presents the detailed statistics of \ourbench{}.
\begin{table}[htb]
    \centering
    \caption{Statistics of \ourbench{}}\label{table:benchmark}
    \footnotesize
    \begin{adjustbox}{width=1\linewidth}
        \begin{tabular}{c|c|c|c|c|c}
        \hline
        \textbf{Type} & \textbf{Analyzer} & \textbf{Kernel} &
        \textbf{OpenSSL} & \textbf{Libav} &
        \textbf{Total}  \\ \hline

        \multirow{3}{*}{NPD}& CodeQL &  70 (22) & 50 (8) & 57 (2) & 177 (32) \\
        & CppCheck & 83 (8)  & 5 (0) & 10 (2) & 98 (10) \\
        & Infer & / & 16 (3) & 6 (0) & 24 (3) \\\hline 
        
        \multirow{1}{*}{UAF}& CodeQL & 16 (0) & 9 (0) & / & 25 (0)  \\\hline 

        \multirow{1}{*}{BOF}& CodeQL & /  & 42 (0) & / & 42 (0) \\\hline 

        \end{tabular}
    \end{adjustbox}
\end{table}

The identifiers used for each static analyzer across three CWEs are provided in Table~\ref{table:command}. Infer and CPPcheck did not detect any warnings for BOF and UAF in the target projects.
\begin{table}[htb]
    \centering
    \caption{Analyzer identifiers acrocss CWEs}\label{table:command}
    \footnotesize
    \begin{adjustbox}{width=0.75\linewidth}
        \begin{tabular}{c|c|c}
        \hline
        \textbf{Type} & \textbf{Analyzer} & \textbf{Command}  \\ \hline

        \multirow{3}{*}{NPD}& CodeQL & MissingNullTest.ql; 
        \\
        & CppCheck & nullPointer;\\
        & Infer & Null Dereference; \\\hline 
        
        \multirow{1}{*}{UAF}& CodeQL &  UseAfterFree.ql
        \\\hline 

        \multirow{1}{*}{BOF}& CodeQL & OverflowBuffer.ql 
        \\\hline 

        \end{tabular}
    \end{adjustbox}
\end{table}

\textbf{Labeling process.} For each warning, two participants will individually label it as True or False; for the inconsistent cases, a third judge will be introduced to solve the conflicts. If the participants cannot reach agreement after discussion, the warning would be filtered out. All the participants have over five-year experience on C/C++; and each would received 200 dollars for the labeling. The experiments are approved by Ethics Review Board.

\section{Baseline Settings}
\label{ap:baseline}

In RQ1, We compare \app{} against two baselines: LLM4SA~\cite{llm4sa} and LLMDFA~\cite{llmdfa}. 
For LLM4SA, we use its public implementation with the following prompt:

\begin{mdframed}
[linecolor=myblue!50,linewidth=2pt,roundcorner=10pt,backgroundcolor=myyellow!20]
\small

\hspace{3.5mm} You are an expert C/C++ programmer.

\textbf{\# Task Description:} 

The code snippet and the bug report will be provided to you for the purpose of examining the presence of the bug within the code snippet. Initially, you need to explain the behavior of the code. Subsequently, you can determine whether the bug is a true positive or a false positive based on the explanation.  To conclude your answer, you will provide one of the following labels: @@@ real bug @@@, @@@ false alarm @@@, or @@@ unknown @@@.

\textbf{\# Bug Report:} \textit{[Bug Report]}

\textbf{\# Code Snippet:} \textit{[Code Snippet]}

\end{mdframed}

For LLMDFA, we omit the Source/Sink Extraction phase and instead use the sources and sinks from \ourbench{} for Dataflow Summarization and Path Feasibility Validation. The prompt template in NPD detection is as follows.

Prompt for Dataflow Summarization:

\begin{mdframed}
[linecolor=myblue!50,linewidth=2pt,roundcorner=10pt,backgroundcolor=myyellow!20]
\small

\hspace{3.5mm}
\textbf{System Role}: You are a C programmer and very good at analyzing C code. 

\textbf{Task Description}: Determine whether two pointer variables at two lines have the same value (i.e., both pointing to the same address or both NULL).

\textbf{Analysis Rules}: 

If they are the same pointer variable and not overwritten between two lines, the answer should be yes.

If the pointer variable a is assigned to the variable b, and there is no overwrite upon them between two lines, the answer should be yes.

[Other Rules]

\textbf{Examples}:

Example 1: User: [Program] [Question]

System: [Answer: Yes][Explanation: y is assigned with x at line 2 and not overwritten between line 2 and line 3, so the value of y before line 3 and x after line 1 are the same]

[Other Examples]

\textbf{Question}: Now I give you a function: [FUNCTION]

Please answer: Dose [VAR1] used at line [LINE1] have the same value as [VAR2] defined at line [LINE2]

Please think it step by step. Return Yes/No with the explanation.

\end{mdframed}

Prompt for Dataflow Validation:

\begin{mdframed}
[linecolor=myblue!50,linewidth=2pt,roundcorner=10pt,backgroundcolor=myyellow!20]
\small

\hspace{3.5mm} \textbf{System Role}: You are a good programmer and familiar with Z3 Python binding.

\textbf{Description}: Please write a Python program using Z3 Python binding to encode the path condition.

\textbf{Path Info}: Here is the path: [path]. Note that the value of [variable] is 0. 

The branch information is as follows: [branch info].

The early jump information is as follows: [early jump info].

\textbf{Synthesis Task}: Please write a Python script to solve the path condition using Z3 Python binding. You can refer to the skeleton: [skeleton]

\textbf{Fixing Task}: Here is the synthesized result of the last round: [script]. When executing the synthesized script, we encounter the following error: [error message]. Please fix the bug and return a runnable script.

\end{mdframed}
\section{Prompt Design of \app{}}
\label{ap:prompt}

\subsection{Prompt in Context-Aware Symbolic Range Reasoning}

Prompt for variable symbolic range reasoning:

\begin{mdframed}
[linecolor=myblue!50,linewidth=2pt,roundcorner=10pt,backgroundcolor=myyellow!20]
\small

\hspace{3.5mm} \textbf{System Role}: You are an expert C/C++ programmer. 

\textbf{Task Description:} Your goal is to determine the possible value ranges for each variable in [conditional statement] assuming the [P] when [conditional statement]  is executed.

Important:

- Analyze each line sequentially through symbolic execution.

- Consider function semantics based on their names.

- Pay special attention to pointer operations and conditions.

- Track how variable ranges change through the execution path.

After the analysis, conclude your answer. 

\textbf{Code Snippet:} [Code Snippet]

Please provide your analysis in clear steps, explaining how you arrived at the value ranges.

\end{mdframed}

Prompt for contextual analysis:

\begin{mdframed}
[linecolor=myblue!50,linewidth=2pt,roundcorner=10pt,backgroundcolor=myyellow!20]
\small

\hspace{3.5mm} \textbf{System Role}: You are an expert C/C++ programmer. 

\textbf{Task Description:} 
Given the function [function name] with body [function body],
your goal is to determine the possible value ranges for the function return value when [P]. If the value range cannot be directly determined and further investigation in deeper function is required, please use the search\_context tool to retrieve the function body for further analysis. 

Important:

... 

\textbf{Code Snippet:} [Code Snippet]

Please provide your analysis in clear steps, explaining how you arrived at the value ranges.

\end{mdframed}

\subsection{Prompt in SMT Query Script Generation}


Prompt in SMT constraints generation:

\begin{mdframed}
[linecolor=myblue!50,linewidth=2pt,roundcorner=10pt,backgroundcolor=myyellow!20]
\small

\hspace{3.5mm} \textbf{System Role}: You are an SMT constraint expert. 

\textbf{Task Description:} 
Convert the following C/C++ conditional expressions into Z3 SMT solver constraints: 
[Conditional Expressions]

When generating Z3 SMT solver scripts, follow these critical guidelines:

[SMT Query Script Generation Rules]

Return only the executable Z3 constraints in the following format:

\verb|```|\textit{python} 

\textit{var\_name1 = Int(`var 1')}

\textit{var\_name2 = Bool(`var 2')}

\textit{s.add(constraint 1)}

\textit{s.add(constraint 2)} 

...

\verb|```|

\end{mdframed}

Prompt in script merging:

\begin{mdframed}
[linecolor=myblue!50,linewidth=2pt,roundcorner=10pt,backgroundcolor=myyellow!20]
\small

\hspace{3.5mm} \textbf{System Role}: You are an expert Z3 constraint solver. 

\textbf{Task Description:} 
I'll provide a Z3 constraint script and a set of additional constraints. Please combine them into a single script and keeping all existing constraints (both from the original script and the new constraint sets). If any new constraints are missing variable definitions, make sure to add or modify the necessary declarations.
Return your solution as a function named \textit{check\_constraints()} that returns a boolean indicating whether the constraints are satisfiable. The function should be ready to execute and no imports or extra content needed.

\textbf{Original Z3 script:} [SMT Query Script]

\textbf{Additional Constraint Set:} [SMT Format Constraints]

\end{mdframed}

Prompt in script fix:

\begin{mdframed}
[linecolor=myblue!50,linewidth=2pt,roundcorner=10pt,backgroundcolor=myyellow!20]
\small

\hspace{3.5mm} \textbf{Task Description:} 
The Z3 script you generated contains errors. Please fix and regenerate the Z3 Python script.  The script should be directly executable.

\textbf{Error details:} [Error Message]

\textbf{Additional Requirements:} [SMT Query Script Generation Rules]

\end{mdframed}

\end{document}